\documentclass[12pt]{article}
\usepackage{fleqn,glas}

\usepackage{graphicx}
\usepackage{epsfig}
\usepackage[figuresright]{rotating}

\newcommand{\AmS}{{\protect\the\textfont2
  A\kern-.1667em\lower.5ex\hbox{M}\kern-.125emS}}

\begin{document}
\begin{titlepage}{GLAS-PPE/1999--19}{November 1999}

\title{Mass Measurement of the W-boson using the ALEPH Detector at LEP}

\author{A.S.Thompson}
       

\begin{abstract}
The W-boson mass has been measured using the ALEPH
detector at LEP. Preliminary results from data taken in 1998 are added
to previous measurements to give\linebreak 
$m_{W}$~=~80.411~$\pm$~0.064(stat.)~$\pm$~0.037(syst.)~$\pm$~0.022(BE-CR)~$\pm$~0.018(LEP)~GeV/$c^{2}$.

\end{abstract}

\vfill
\conference{Presented at the \\
8th International Conference on Hadron Spectroscopy, IHEP, Beijing, \\ 
August 24-28 1999}

\end{titlepage}

\section{INTRODUCTION}

In this paper we discuss the measurement of the W mass from direct 
reconstruction of the invariant mass of the decay products in the channels
WW$\rightarrow$q\={q}q\={q} (4q) and WW$\rightarrow\ell\nu$q\={q}.
Preliminary results are presented for data collected in ALEPH\cite{aleph}
during 1998 with an integrated luminosity of 174.2${pb^{-1}}$ at 188.63GeV.

\section{EVENT SELECTIONS}

\subsection{WW $\rightarrow$ q\={q}q\={q} events}

Events are preselected to remove radiative returns to the Z and clustered into
four jets using the \verb+DURHAM-PE+ algorithm, defining $y_{cut}>0.001$. 
Events are vetoed if a charged track in a jet carries more than 90\% of the jet
energy or if there is more than 95\% of electromagnetic energy in a $1^{\circ}$
cone around any particle.

A Neural Network with fourteen input variables (NN14) is used to perform 
the final selection. Training is performed with an independent sample
of the standard Monte-Carlo (\verb+KORALW+\cite{Skrz96} hadronised with 
\verb+JETSET+\cite{Sjos94}) and comparable samples of q\={q}$(\gamma)$
and ZZ events, both generated using \verb+PYTHIA+\cite{Sjos94}, to simulate the
background. Events with NN$14>0.3$ are used to extract the  
W mass\cite{aleph189}.

\subsection{WW $\rightarrow$ e$\nu$q\={q} and 
WW $\rightarrow \mu\nu$q\={q} events}

The total charged energy and multiplicity are used to preselect events
with a further cut on the total longitudinal momentum and visible energy to
remove radiative returns to the Z. The lepton candidates are chosen to be
more energetic and isolated than the other charged tracks and are 
identified as an electron or muon in the detector. The energy of electron
candidates are corrected for possible bremsstrahlung photons detected in
the electromagnetic calorimeter.

The \verb+DURHAM-PE+ algorithm is used to force two jets from objects not 
used to reconstruct the lepton defining $y_{cut}>0.0003$. A probability 
for an event to come from a signal process is determined from Monte Carlo 
reference samples using the lepton energy and isolation and the event total
transverse momentum. Selected events are required to have a probability
greater than 0.4.

\subsection{WW $\rightarrow \tau\nu$q\={q} events} 

A similar preselection to that used for electron and muon events is applied
with additional constraints to remove 
events with i) energy around the beam line, ii) isolated, energetic
photons and iii) those already selected as electron or muon candidate
events. 

A tau jet is constructed from one or three charged tracks and two other jets 
are forced using the \verb+JADE+ algorithm. The tau jet must be that jet
most anti-parallel to the missing momentum vector and isolated from
the other jets. As with electron and muon candidate events, a probability is
calculated and the same cut applied.

\section{EXTRACTION OF THE W-MASS}

\subsection{WW $\rightarrow$ q\={q}q\={q} events}

To improve the invariant mass resolution a four constraint kinematic
fit is applied to the four jets to conserve energy (as provided by LEP)
and momention. Corrections are applied to the jets to take into account
particle losses in the detector. The masses from the three combinations
of di-jets formed from the fitted jets are rescaled according to
$m_{ij}^{resc}/m_{ij}=E_{beam}/(E_{i}+E_{j})$, where $E_{i}$ and $E_{j}$ are 
the measured jet energies.

A pairing algorithm is applied to the di-jet combinations to select that
which most closely corresponds to a WW pair. The combination with
the smallest mass difference between rescaled masses is chosen provided
one mass lies in the window 60 to 86 GeV/$c^{2}$ and the other 
74 to 86 GeV/$c^{2}$.

A binned Monte-Carlo reweighting procedure is employed to find the
value of $m_{W}$ which best fits the mass distributions. Events from a 
\verb+KORALW+\cite{Skrz96} sample with equivalent background are 
reweighted with a CC03 matrix element to provide a two-dimensional 
probability density function for the minimisation using a single parameter,
$m_{W}$. Variable binning controlled by the density of Monte-Carlo is employed
optimised to produce a stable result. The W-width is allowed to vary with
$m_{W}$ according to the Standard Model.

In order to check that the procedure does not introduce any biases, 
Monte Carlo samples are generated with $m_{W}$ in the range 79.35 and 81.35
GeV/$c^{2}$. Treating these samples as data, no significant offsets 
in the masses measured are found.

\subsection{WW $\rightarrow \ell\nu$q\={q} events} 

A two constraint fit\cite{aleph183} is applied to each event by minimising  a 
$\chi^{2}$ 
constructed from the deviations of selected parameters of the 
jets and leptons from their true values and demanding that the hadron and 
lepton invariant masses are equal. The single fitted mass obtained for 
each event must lie in the window 74 to 94.5 GeV/$c^{2}$.

A reweighting procedure similar to that employed in the four quark channel
is used to fit the mass distribution. Fixed binning is used for the electron 
and muon channels and variable binning is retained for the tau channel.

\begin{table}[h]
\small
\renewcommand{\baselinestretch}{0.8}
\caption{Summary of Systematic Errors on the measurement of the W-mass 
(MeV/$c^{2}$)}
\label{table:1}
\newcommand{\m}{\hphantom{$0$}}
\newcommand{\cc}[1]{\multicolumn{1}{c}{#1}}
\renewcommand{\tabcolsep}{2pc} 
\renewcommand{\arraystretch}{1.0} 
\begin{tabular}{@{}p{5.0cm}lllll}
\hline
Source                    & 4q      & e       & $\mu$   & $\tau$  \\
\hline \hline
\multicolumn{5}{l}{Errors correlated between channels} \\
\hline
Calorimeter calibrations  &   30 &   27 &   14 &   19 \\
Charged particle tracking &      & \m7  & \m3  & \m3  \\
Jet corrections           & \m8  &   14 & \m4  & \m7  \\
Parton fragmentation      &   35 &   25 &   25 &   30 \\
Initial State radiation   &   10 & \m5  & \m5  & \m5  \\
LEP energy                &   17 &   17 &   17 &   17 \\
\hline
\multicolumn{5}{l}{Uncorrelated Errors} \\  
\hline 
Reference MC statistics   & 10   & 16   & 15   & 23   \\
Background contamination  & 10   & \m8  & \m1  & 25   \\
Colour reconnection       & 25   &      &      &      \\
Bose-Einstein effects     & 50   &      &      &      \\
\hline \hline
Total                     & 77   & 47   & 37   & 53   \\
\hline
\end{tabular}\\[2pt]
\end{table}

\section{SYSTEMATIC UNCERTAINTIES}

The systematic errors summarised in Table 1. 
Uncertainties due to the detector are determined using data taken at
the Z at intervals during the high energy running. Particles not seen
by the detector cause discrepancies in the jet finding, the effect
is estimated by matching jets built from Monte Carlo tracks
before and after passing them through the detector simulation.

The error due to  parton fragmentation effects is measured by determining
the mass shift when \verb+HERWIG+\cite{Marc92} rather than \verb+JETSET+ is 
used for hadronisation. The effect of initial state radiation is estimated
by comparing the use of 1st and second order matrix elements. The LEP
energy error is that given by LEP.

Colour reconnection between parton pairs in the q\={q}q\={q} channel is
studied using variants on \verb+JETSET+ or \verb+HERWIG+. The Bose-Einstein
effect has been studied using Z-peak data\cite{aleph99}, the systematic error 
is determined from the shift in the mass measured when the parameters obtained
at the Z peak are applied to Monte Carlo events using a modified 
\verb+JETSET+. This effect is applied between to W decay products in the
q\={q}q\={q} channel.

\begin{table}[h]
\small
\renewcommand{\baselinestretch}{0.8}
\caption{W mass measurement results at 189 GeV}
\label{table:2}
\newcommand{\m}{\hphantom{$0$}}
\newcommand{\cc}[1]{\multicolumn{1}{c}{#1}}
\renewcommand{\tabcolsep}{2pc} 
\renewcommand{\arraystretch}{1.0} 
\begin{tabular}{@{}p{3.0cm}lllll}
\hline
                              &$m_{W}$ GeV/$c^{2}$& stat. & stat. & BE-CR  \\
\hline \hline
Hadronic (4q)                 & 80.561            & 0.116 & 0.053 & 0.056  \\
\hline \hline
WW $\rightarrow e\nu$q\={q}   & 80.524            & 0.180  & 0.047 &    \\
WW $\rightarrow \mu\nu$q\={q} & 80.297            & 0.164  & 0.037 &    \\
WW $\rightarrow \tau\nu$q\={q}& 80.461            & 0.332  & 0.053 &    \\
\hline
Semileptonic                  & 80.406            & 0.114  & 0.033 &    \\
\hline \hline
Weighted Mean                 & 80.472            & 0.081  & 0.043 & 0.025 \\
\hline
\end{tabular}\\[2pt]
\end{table}

\begin{figure}[h]
\begin{center}
\psfig{file=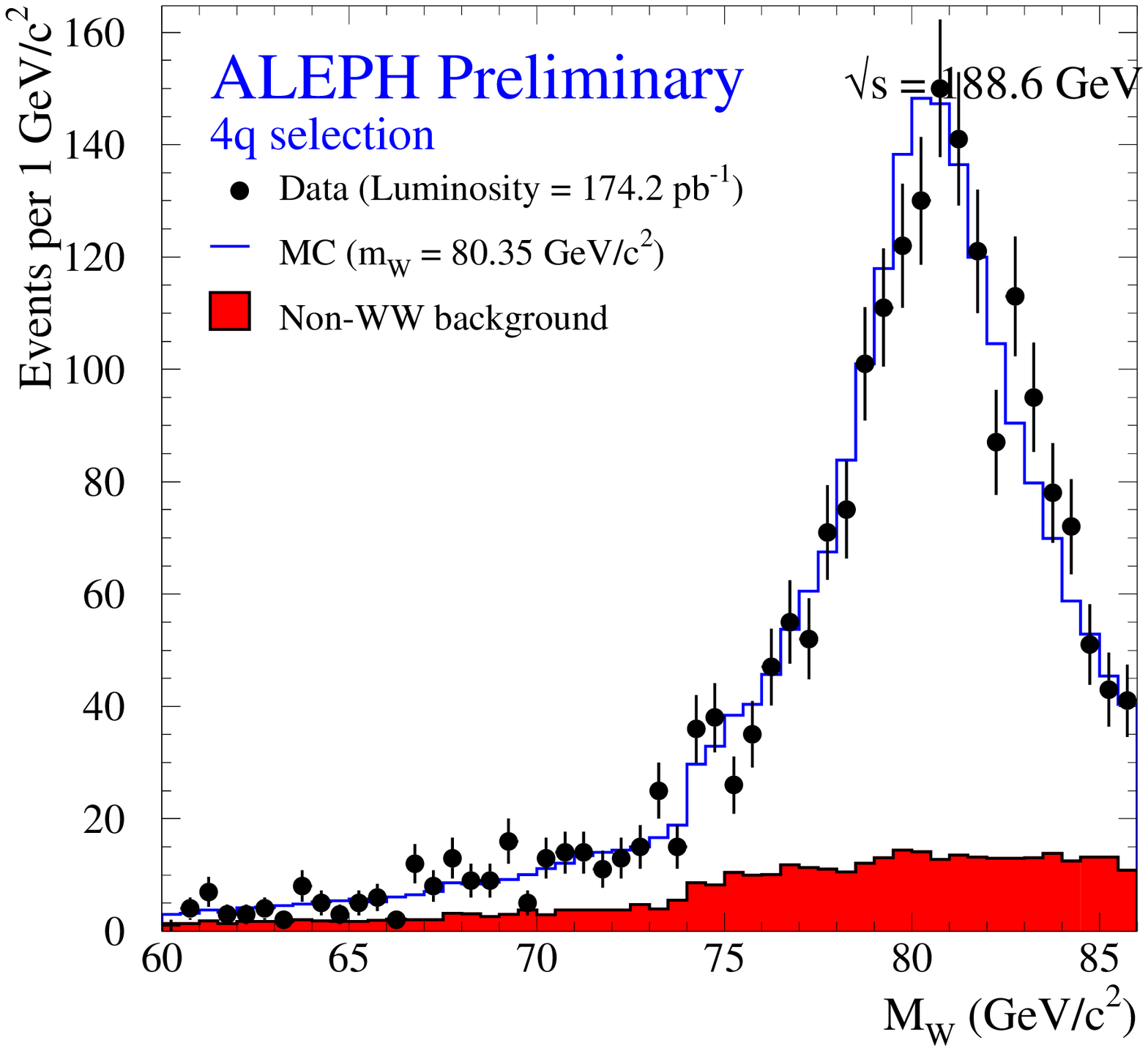,width=64mm}
\psfig{file=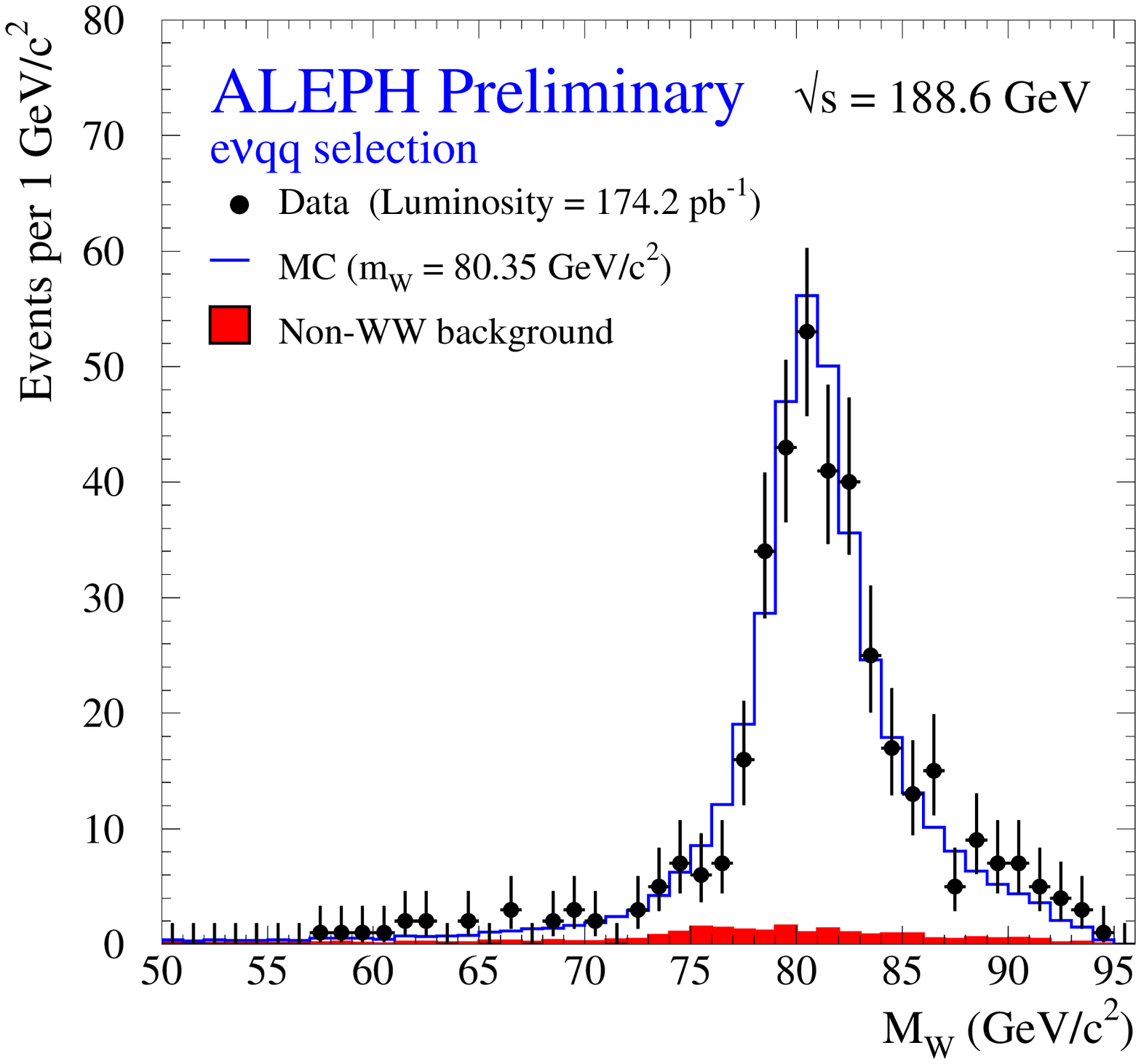,width=67mm}
\psfig{file=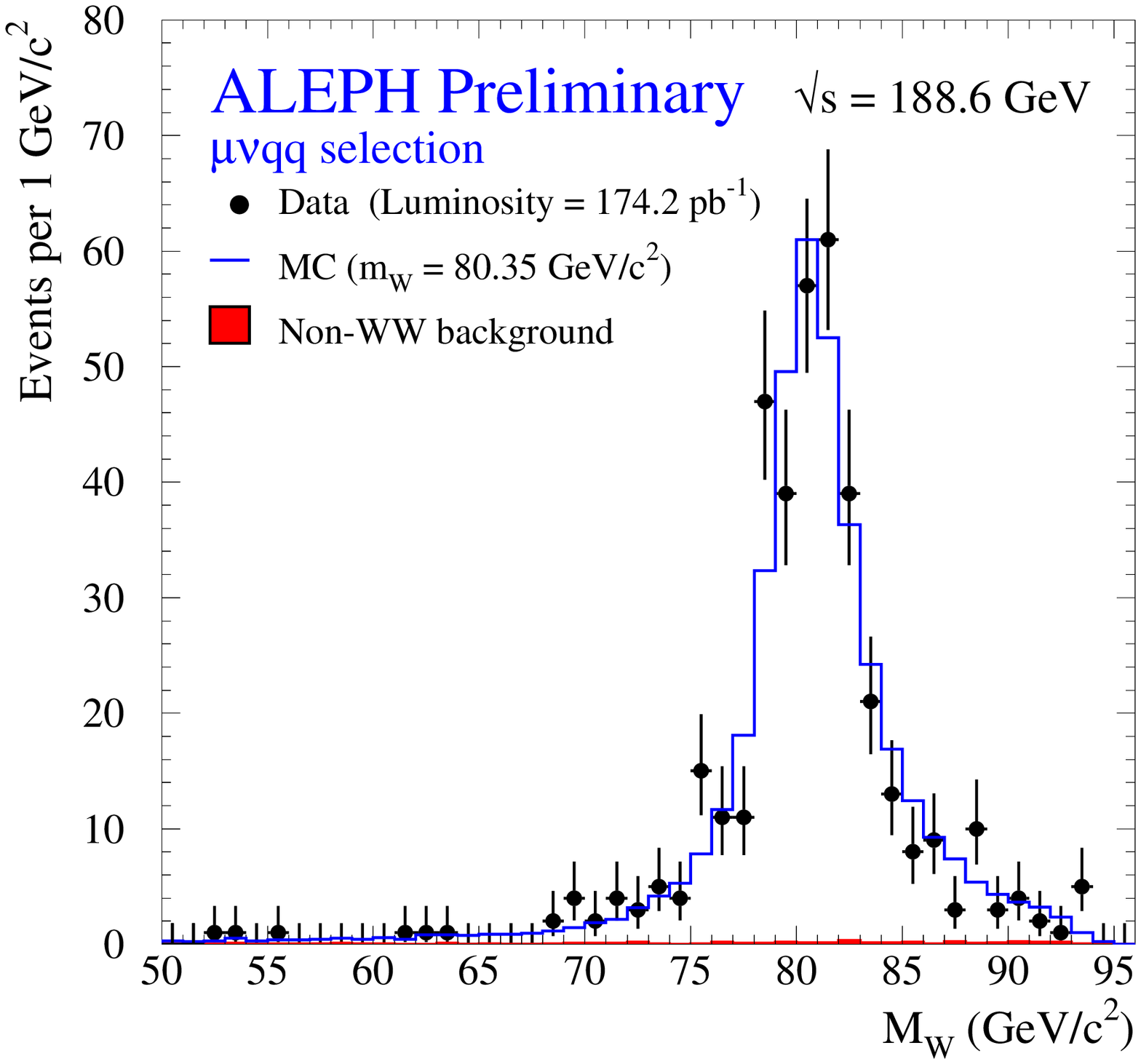,width=67mm}
\psfig{file=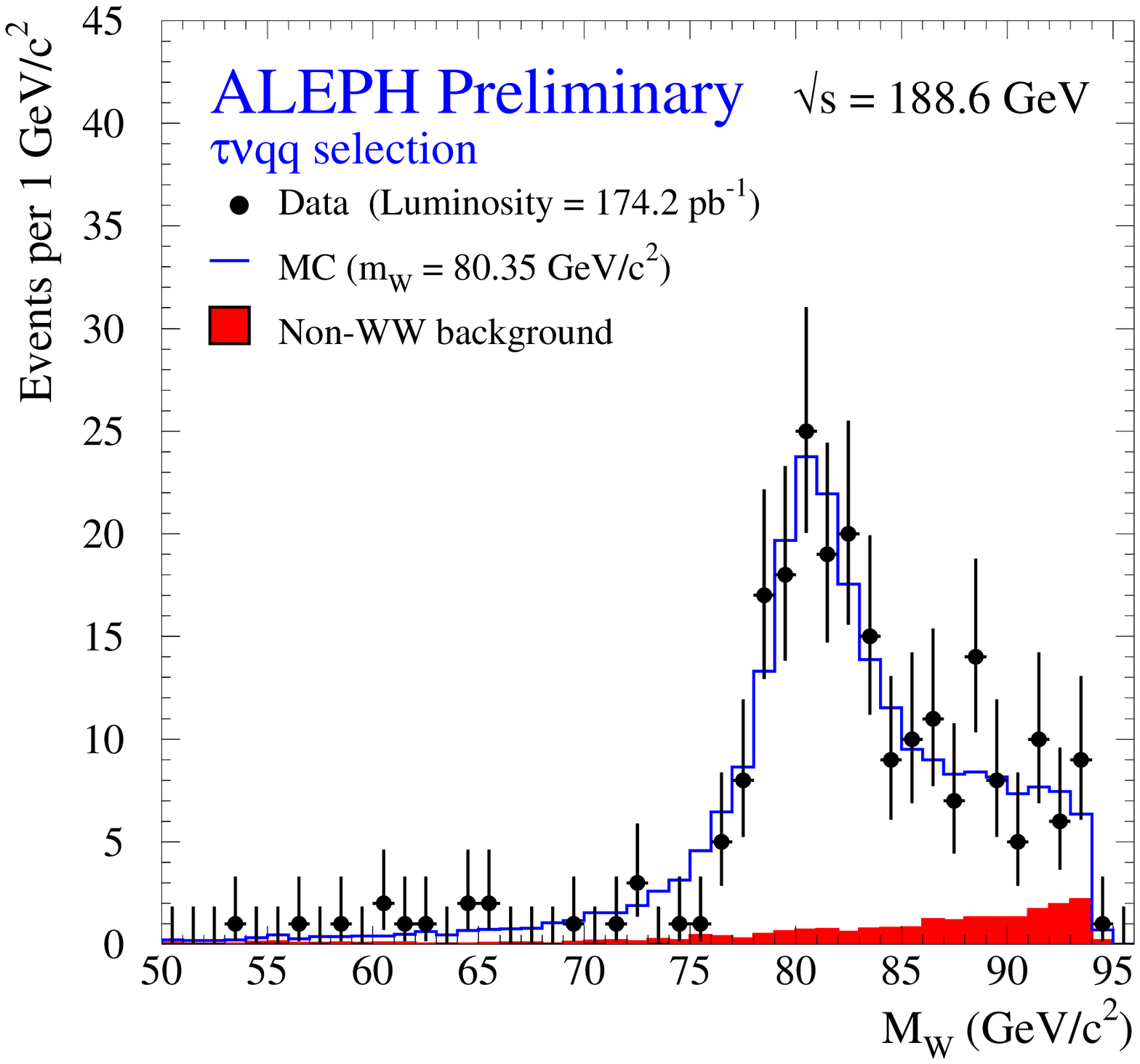,width=67mm}
\caption{Mass distributions for the 4q (rescaled, two entries per event),
e, $\mu$ and $\tau$ channels. The Monte Carlo prediction is for 
$m_{W}$ = 80.35 GeV/$c^{2}$.}
\end{center}
\end{figure}

\section{MASS MEASUREMENT RESULTS}

The mass distributions obtained are shown in figure 1.
The preliminary results for 189 GeV are given in Table 2 for each channel
where the BE-CR error is that for the Bose-Einstein and Colour reconnection
systematics added in quadrature.

Results for data taken at 172 and 183 GeV have been published 
\cite{aleph172,aleph183},
combining these and the results above gives
\newline
$m_{W}^{hadronic}$ = 80.561 $\pm$ 0.095(stat.) $\pm$ 0.050(syst.) $\pm$ 0.056(BE-CR) GeV/$c^{2}$
\newline
$m_{W}^{leptonic~}$ = 80.343 $\pm$ 0.089(stat.) $\pm$ 0.041(syst.) GeV/$c^{2}$.

The current weighted average of all ALEPH results is obtained by including  measurements
from W pair cross sections at 161 and 172 GeV \cite{aleph161}, giving 

$m_{W}$ = 80.411 $\pm$ 0.064(stat.) $\pm$ 0.037(syst.) $\pm$ 0.022(BE-CR) 
$\pm$ 0.018(LEP) GeV/$c^{2}$

\end{document}